\documentclass[prl,aps,showpacs,twocolumn,twoside,superscriptaddress]{revtex4}

\usepackage{amsmath,amsfonts,amssymb,color,epsfig,graphics,graphicx,latexsym,revsymb,theorem,verbatim}

\newtheorem{definition}{Definition}
\newtheorem{proposition}[definition]{Proposition}
\newtheorem{lemma}[definition]{Lemma}

\newtheorem{theorem}[definition]{Theorem}
\newtheorem{corollary}[definition]{Corollary}
\newtheorem{conjecture}[definition]{Conjecture}

\newtheorem{remark}[definition]{Remark}

\def\squareforqed{\hbox{\rlap{$\sqcap$}$\sqcup$}}
\def\qed{\ifmmode\squareforqed\else{\unskip\nobreak\hfil
\penalty50\hskip1em\null\nobreak\hfil\squareforqed
\parfillskip=0pt\finalhyphendemerits=0\endgraf}\fi}
\def\endenv{\ifmmode\;\else{\unskip\nobreak\hfil
\penalty50\hskip1em\null\nobreak\hfil\;
\parfillskip=0pt\finalhyphendemerits=0\endgraf}\fi}
\newenvironment{proof}{\noindent \textbf{{Proof.~} }}{\qed}

\def\bcj{\begin{conjecture}}
\def\ecj{\end{conjecture}}
\def\bcr{\begin{corollary}}
\def\ecr{\end{corollary}}
\def\bd{\begin{definition}}
\def\ed{\end{definition}}
\def\bea{\begin{eqnarray}}
\def\eea{\end{eqnarray}}
\def\bem{\begin{enumerate}}
\def\eem{\end{enumerate}}
\def\bim{\begin{itemize}}
\def\eim{\end{itemize}}
\def\bl{\begin{lemma}}
\def\el{\end{lemma}}
\def\bpf{\begin{proof}}
\def\epf{\end{proof}}
\def\bpp{\begin{proposition}}
\def\epp{\end{proposition}}
\def\br{\begin{remark}}
\def\er{\end{remark}}
\def\bt{\begin{theorem}}
\def\et{\end{theorem}}

\def\r{\rho}
\def\s{\sigma}

\def\ph{\varphi}

\def\ps{\psi}

\def\Ps{\Psi}

\newcommand{\nc}{\newcommand}

\nc{\cA}{{\cal A}} \nc{\cB}{{\cal B}} \nc{\cC}{{\cal C}}
\nc{\cD}{{\cal D}} \nc{\cE}{{\cal E}} \nc{\cF}{{\cal F}}
\nc{\cG}{{\cal G}} \nc{\cH}{{\cal H}} \nc{\cI}{{\cal I}}
\nc{\cJ}{{\cal J}} \nc{\cK}{{\cal K}} \nc{\cL}{{\cal L}}
\nc{\cM}{{\cal M}} \nc{\cN}{{\cal N}} \nc{\cO}{{\cal O}}
\nc{\cP}{{\cal P}} \nc{\cR}{{\cal R}} \nc{\cS}{{\cal S}}
\nc{\cT}{{\cal T}} \nc{\cV}{{\cal V}} \nc{\cX}{{\cal X}}
\nc{\cZ}{{\cal Z}}

\def\Tr{\mathop{\rm Tr}\nolimits}

\def\max{\mathop{\rm max}}

\def\rank{\mathop{\rm rank}}
\def\rk{\mathop{\rm rk}}

\def\tr{\mathop{\rm Tr}}

\def\bigox{\bigotimes}

\def\ox{\otimes}

\newcommand{\ket}[1]{|#1\rangle}
\newcommand{\proj}[1]{| #1\rangle\!\langle #1 |}

\newcommand{\cmp}{Comm. Math. Phys.}

\newcommand{\pla}{Phys. Lett. A}

\begin{document}
\title{Weaker entanglement guarantees stronger entanglement}

\author{Masahito Hayashi}
\email{hayashi@math.is.tohoku.ac.jp}
\address{Graduate School of Information Sciences, Tohoku
University, Aoba-ku, Sendai, 980-8579, Japan}
\address{Centre for Quantum Technologies, National University of Singapore, 3 Science Drive 2, 117542, Singapore}
\author{Lin Chen$^2$}
\email{cqtcl@nus.edu.sg(Corresponding~Author)}

\begin{abstract}
The monogamy of entanglement is one of the basic quantum mechanical
features, which says that when two partners Alice and Bob are more
entangled then either of them has to be less entangled with the
third party. Here we qualitatively present the converse monogamy of
entanglement: given a tripartite pure system and when Alice and Bob
are weakly entangled, then either of them is generally strongly
entangled with the third party. Our result leads to the
classification of tripartite pure states based on bipartite reduced
density operators, which is a novel and effective way to this
long-standing problem compared to the means by stochastic local
operations and classical communications. We also systematically
indicate the structure of the classified states and generate them.

\end{abstract}

\date{ \today }




\maketitle

\noindent \textit{Introduction}. The monogamy of entanglement is a
purely quantum phenomenon in physics \cite{ckw00} and has been used
in various applications, such as bell inequalities \cite{sg01} and
quantum security \cite{KW}. In general, it indicates that the more
entangled the composite system of two partners Alice $(A)$ and Bob
$(B)$ is, the less entanglement between $A$ $(B)$ and the
environment $E$ there is. The security of many quantum secret
protocols can be guaranteed quantitatively \cite{KW,TH}. However the
converse statement generally doesn't hold, namely when $A$ and $B$
are less entangled, we cannot decide whether $A$ $(B)$ and $E$ are
more entangled. In fact even when the formers are classically
correlated namely separable \cite{werner89}, the latters may be also
separable. For example, this is realizable by the tripartite
Greenberger-Horne-Zeilinger (GHZ) state.

Nevertheless, it is still important to {\em qualitatively}
characterize the above converse statement in the light of the
hierarchy of entanglement of bipartite systems. Such a
characterization defines a converse monogamy of entanglement, and
there is no classical counterpart. Besides, it is also expected to
be helpful for treating a quantum multi-party protocol when the
third party helps the remaining two parties, for it guarantees the
property of one reduced density operator from another. To justify
the hierarchy of entanglement, we recall six well-known conditions,
i.e., the separability, positive-partial-transpose (PPT)
\cite{hhh96,HHH}, non-distillability of entanglement under local
operations and classical communications (LOCC) \cite{HHH}, reduction
property (states satisfying reduction criterion) \cite{HH},
majorization property \cite{Hiro} and negativity of conditional
entropy \cite{HH}. These conditions form a hierarchy since a
bipartite state satisfying the former condition will satisfy the
latter too. Therefore, the strength of entanglement in the states
satisfying the conditions in turn becomes gradually {\em weak}.

For example, the hierarchy is closely related to the distilability
of entanglement \cite{HHH}. While PPT entangled states cannot be
distilled to Bell states for implementing quantum information tasks,
Horodecki's protocol can distill a state that violates reduction
criterion \cite{HH}. That is, the former entangled state is useless
as a resource while the latter entangled state is useful. So the
usefulness of entangled states can be characterized by this
hierarchy.

\begin{figure}
  \includegraphics[width=3cm]{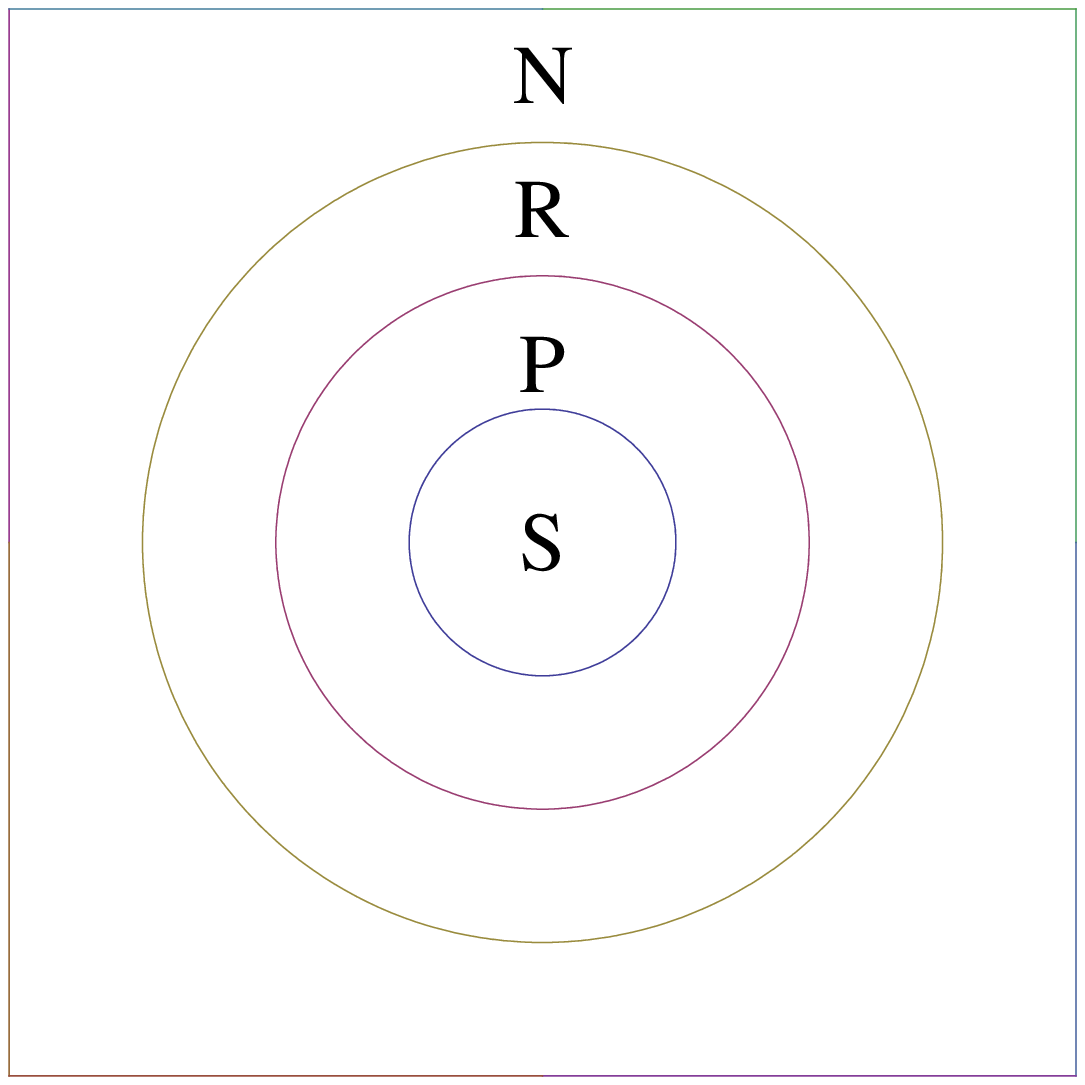}\\
  \caption{\label{fig:SPRN} Hierarchy of bipartite states in terms of four sets S, P, R, N.
  Intuitively, the sets S and P form all
  PPT states, the sets S, P, R and N form all states satisfying and violating
  the reduction criterion, respectively. So the four sets
  constitute the set of bipartite states and there is no intersection between any two sets. The strength of
  entanglement of the four sets becomes weak in turn, i.e., $S \le P \le R  \le
N$.}
\end{figure}

In this Letter, for simplicity we consider three most important
conditions, namely the separability, PPT and the reduction criteria.
Then we establish a hierarchy of entanglement consisting of four
sets: separable states (S), non-separable PPT states (P), non-PPT
reduction states (R), and non-reduction states (N), see Figure
\ref{fig:SPRN}. We show that when the entangled state between $A$
and $B$, i.e., $\r_{AB}$ belongs to the set $R$, then the state
between $A$ $(B)$ and $E$, i.e., $\r_{AE} (\r_{BE})$  belongs to the
set $R$ or $N$. Similarly when $\r_{AB}$ belongs to $P$, then
$\r_{AE} (\r_{BE})$ belongs to $N$. Hence we can qualitatively
characterize the converse monogamy of entanglement as follows: when
the state $\r_{AB}$ is weakly entangled, then $\r_{AE}$ is generally
strongly entangled in terms of the four cases $S,P,R,N$. These
assertions follow from a corollary of Theorem
\ref{thm:sixconditions} to be proved later.
\begin{theorem}
\label{thm1} Suppose the tripartite state $\ket{\Ps}_{ABC}$ has a
PPT reduced state $\rho_{BC}$. Then the reduced state $\rho_{AB}$ is
separable if and only if it satisfies the reduction criterion.
\end{theorem}
From this theorem, we will solve two conjectures on the existence of
specified tripartite state proposed by Thapliyal in 1999
\cite{thapliyal99}. On the other hand, the theorem also helps
develop the classification of tripartite pure states based on the
three reduced states, each of which could be in one of the four
cases $S,P,R,N$. So there are at most $4^3=64$ different kinds of
tripartite states. Evidently, some of them do not exist due to
Theorem \ref{thm1}. It manifests that the quantum behavior of a
global system is strongly restricted by local systems. By
generalizing to many-body systems, we can realize the quantum nature
on macroscopic size in terms of the microscopic physical systems.
This is helpful to the development of matter and material physics
\cite{hw11}. Hence, in theory it becomes important to totally
identify different tripartite states.

To explore the problem, we describe the properties for reduced
states $\rho_{AB}$, $\rho_{BC}$, and $\rho_{CA}$ of the state
$|\Psi\rangle$ by $X_{AB},X_{BC},X_{CA}$ that take values in
$S,P,R,N$. The subset of such states $|\Psi\rangle$ is denoted by
$\cS_{X_{AB}X_{BC}X_{CA}}$, and the subset is non-empty when there
exists a tripartite state in it. For example, the GHZ state belongs
to the subset $\cS_{SSS}$. Furthermore as is later shown in Table
\ref{tab:tripartite}, $|\Psi\rangle$ belongs to the subset
$\cS_{SSN}$ when the reduced state $\rho_{CA}$ is an entangled
maximally correlated state \cite{rains99}. By Theorem \ref{thm1},
one can readily see that any non-empty subset is limited in eight
essential subsets
$\cS_{SSS},\cS_{SSN},\cS_{SNN},\cS_{PNN},\cS_{RRR},\cS_{RRN},\cS_{RNN},\cS_{NNN}$.
Hence up to permutation, the number of non-empty subsets for
tripartite pure states is at most $18=1\times 3+3\times 5$. We will
demonstrate that the 18 subsets are indeed non-empty by
explicit examples. These subsets are not preserved under the
conventional classification by the invertible stochastic LOCC
(SLOCC) \cite{dvc00,cds08,ccd10}.

Furthermore, we show that the subsets form a commutative monoid and
it is a basic algebraic concept. We systematically characterize the
relation of the subsets by generating them under the rule of monoid.


\medskip
\noindent \textit{Unification of entanglement criterion}. In quantum
information, the following six criteria are extensively useful for
studying bipartite states $\r_{AB}$ in the space $\cH_A \ox \cH_B$.
\begin{description}
\item[(1)]
Separability: $\rho_{AB}$ is the convex sum of product states
\cite{werner89}.

\item[(2)]
PPT condition: the partial transpose of $\rho_{AB}$ is semidefinite
positive \cite{hhh96}.

\item[(3)]
Non-distillability: no pure entanglement can be asymptotically
extracted from $\rho_{AB}$ under LOCC, no matter how many copies are
available \cite{HHH}.

\item[(4)]
Reduction criterion: $\rho_{A}\otimes I_B \ge \rho_{AB}$ and $I_A
\otimes \rho_{B} \ge \rho_{AB}$ \cite{HH}.

\item[(5)]
Majorization criterion: $\rho_{A} \succ \rho_{AB}$ and $\rho_{B}
\succ \rho_{AB}$ \cite{Hiro}.

\item[(6)]
Conditional entropy criterion: $H_{\rho}(B|A)=  H(\rho_{AB})-
H(\rho_{A}) \ge 0$ and $H_{\rho}(A|B)=  H(\rho_{AB})- H(\rho_{B})
\ge 0$, where $H$ is the von Neumann entropy.
\end{description}
It is well-known that the relation $ {\bf (1)} \Rightarrow {\bf (2)}
\Rightarrow {\bf (3)} \Rightarrow {\bf (4)} \Rightarrow {\bf (5)}
\Rightarrow {\bf (6)}$ holds for any state $\r_{AB}$
\cite{HHH,HH,Hiro}. In particular apart from ${\bf (2)} \Rightarrow
{\bf (3)}$ whose converse is a famous open problem \cite{dss00}, all
other relations are strict. We will show that these conditions
become equal when we further require $\rho_{BC}$ is PPT. First,
under this restriction the conditions {\bf (5)} and {\bf (6)} are
respectively simplified into {\bf(5')} $\rho_{A} \sim \rho_{AB}$ and
{\bf(6')} $H(\rho_{A}) = H(\rho_{AB})$. Second, when $\rho_{BC}$ is
PPT, since $\rho_{AB} \succ \rho_{A}$ holds, the above two
conditions {\bf (5')} and {\bf (6')} are equivalent. Now we have

\begin{theorem}
\label{thm:sixconditions}
 For a tripartite state $\ket{\Ps}_{ABC}$ with
 a PPT reduced state $\rho_{BC}$, conditions {\bf (1)}-{\bf (6)},
 {\bf (5')}, and {\bf (6')} are equivalent for $\rho_{AB}$.
\end{theorem}
The proof is given in the end of this paper. We can readily get
Theorem \ref{thm1} from Theorem \ref{thm:sixconditions}, and provide
its operational meaning as the main result of the work.
 \bt
 \label{thm:conversemonogamy}
 $\bf{(Converse~monogamy~of~entanglement).}$ Consider a tripartite
 state $\ket{\Ps}_{ABE}$ with entangled reduced states $\r_{AB}, \r_{AE}$ and
 $\r_{BE}$. When $\r_{AB}$ is a weakly entangled state $P$
 ($R$), the states $\r_{AE}$ and $\r_{BE}$ are strongly
 entangled states $N$ ($R$ or $N$).
 \et
To our knowledge, the converse monogamy of entanglement is another
basic feature of quantum mechanics and there is no classical
counterpart since classical correlation can only be "quantified". In
contrast, quantum entanglement has qualitatively different levels of
strength and they have essentially different usefulness from each
other. For example the states in the subset $P$ cannot be distilled
while those in $N$ are known to be distillable \cite{HH}. So only
the latter can directly serve as an available resource for quantum
information processing and it implies the following paradox. {\bf A
useless entangled state between $A$ and $B$ strengthens the
usefulness of entanglement resource between $A$ ($B$) and the
environment}. Therefore, the converse monogamy of entanglement
indicates a dual property to the monogamy of entanglement: Not only
the amount of entanglement, the usefulness of entanglement in a
composite system is also restricted by each other.

Apart from bringing about the converse monogamy of entanglement,
Theorem \ref{thm:sixconditions} also promotes the study over a few
important problems. For instance, the equivalence of {\bf (1)} and
{\bf (2)} is a necessary and sufficient condition of deciding
separable states, beyond that for states of rank not exceeding 4
\cite{hhh96,hlv00,cd11}. Besides, the equivalence of {\bf (2)} and
{\bf (3)} indicates another kind of non-PPT entanglement activation
by PPT entanglement \cite{evw01}. For later convenience, we
explicitly work out the expressions of states satisfying the
assumptions in Theorem \ref{thm:sixconditions}.
 \bl
 \label{le:SNSSSS}¡¡
 The tripartite pure state with two PPT reduced states,
 namely belongs to the subset $\cS_{SNS}$
 or $\cS_{SSS}$ if and only if it has the form $\sum_i \sqrt{p_i}
 \ket{b_i,i,i}$ up to local unitary operators. In other words, the
 reduced state $\rho_{BC}$ is maximally correlated.
 \el
For the proof see Lemma 11 in \cite{cd11}. We apply our results to
handle two open problems in FIG. 4 of \cite{thapliyal99}, i.e.,
whether there exist tripartite states with two PPT bound entangled
reduced states, and tripartite states with two separable and one
bound entangled reduced states. Here we give negative answers to
these conjectures in terms of Theorem~\ref{thm:sixconditions} and
Lemma \ref{le:SNSSSS}. As the first conjecture is trivial, we
account for the second conjecture. Because the required states have
the form $\sum^d_{i=1} \sqrt{p_i}\ket{ii}\ket{c_i}$, where $\r_{AC}$
and $\r_{BC}$ are separable. So the reduced state $\r_{AB}$ is a
maximally correlated state, which is either separable or
distillable. It readily denies the second conjecture.

\medskip
\noindent \textit{Classification with reduced states}. Theorem
\ref{thm:conversemonogamy} says that the quantum correlation between
two parties of a tripartite system is dependent on the third party.
From Theorem \ref{thm:sixconditions} and the discussion to
conjectures in \cite{thapliyal99}, we can see that the tripartite
pure state with some specified bipartite reduced states may not
exist. This statement leads to a classification of tripartite states
in terms of the three reduced states \cite{sa08}. As a result, we
obtain the different subsets of tripartite states in Table
\ref{tab:tripartite} in terms of tensor rank and local ranks of each
one-party reduced state. In the language of quantum information, the
tensor rank of a multipartite state, also known as the Schmidt
measure of entanglement~\cite{eb01}, is equal to the least number of
product states to expand this state. For instance, any multiqubit
GHZ state has tensor rank two. So tensor rank is bigger or equal to
any local rank of a multipartite pure state. As tensor rank is
invariant under invertible SLOCC \cite{cds08}, it has been widely
applied to classify SLOCC-equivalent multipartite states recently
\cite{ccd10}.

Here we will see that, tensor rank is also essential to the
classification in Table \ref{tab:tripartite}. First the statement
for the subsets $\cS_{SSS},\cS_{SSN}$ and $\cS_{SNN}$ follows from
Lemma \ref{le:SNSSSS}, and Lemma 2 in \cite{dfx07}, respectively.
Next to see the statement for $\cS_{PNN}$, it suffices to recall the
following fact \cite{hlv00}.
\begin{lemma}
  \label{le:rankAB=rankA,B}
  A $M \times N$ state $\r_{AB}$ with rank $N$ is PPT if and only
  if it is separable and is the convex sum of just $N$ product states, i.e.,
  $\r_{AB}=\sum^N_{i=1}
  p_i \proj{a_i,b_i}$.
\end{lemma}
Next, we characterize the subset $\cS_{RNN}$ by the tensor rank of
states in this subset.
\begin{lemma}
\label{ha-l-2} Assume that $\rk(\Ps) = \max \{ d_A, d_B \}$. Then
the conditions {\bf (1)}-{\bf (4)} are equivalent for $\rho_{AB}$.
\end{lemma}
\begin{proof}
It suffices to show that the state $\rho_{AB}$ is separable when it
satisfies the reduction criterion. Let $|\Psi
\rangle=\sum_{i=1}^{\rk(\Ps)} \sqrt{p_i} |a_i,b_i,c_i\rangle$, and
$V_A$ an invertible matrix such that $V_A|i\rangle=|a_i\rangle$.
Now, we focus on the pure state $|\Psi' \rangle= K V_A^{-1} \otimes
I_B\otimes I_C |\Psi \rangle$, where $K$ is the normalized constant.
Then the reduced state $\rho_{AB}'$ satisfies $\rho_A' \otimes I_B
\ge \rho_{AB}'$, and hence $\rho_A' \succ \rho_{AB}'$ \cite{Hiro}.
Since $\rho_{BC}'$ is separable, we have $\rho_A' \sim \rho_{AB}'$.
So the state $\rho_{AB}'$, and equally $ \rho_{AB} $  is separable
in terms of Theorem \ref{thm:sixconditions}.
\end{proof}
It's noticeable that under the same assumption in this lemma, the
equivalence between conditions {\bf (1)}-{\bf (5)} does not hold due
to the counterexample, such as the symmetric state $|\Psi_a\rangle:=
\frac{1}{\sqrt{r+7}}(\sum_{i=2}^r|iii\rangle + (|1\rangle +|2
\rangle )(|1\rangle +|2 \rangle )(|1\rangle +|2 \rangle ))$, which
belongs to the subset ${\cS}_{NNN}$. It indicates that tensor rank
alone is not enough to characterize the hierarchy of bipartite
entanglement.

Besides it follows from the definition of reduction criterion that
when the state $\rho_{AB}$ satisfies condition {\bf (4)}, we have
$d_C \ge d_A, d_B $. This observation and Lemma \ref{ha-l-2} justify
the statement for $\cS_{RNN}$ in table \ref{tab:tripartite}. In
order to show the tightness of these inequalities, we consider the
non-emptyness of the subset $\cS_{RNN}$ with the boundary types
$r = d_C > d_A= d_B$ and $r > d_C = d_A = d_B$. To justify the former type,
it suffices to consider the state $\frac{1}{\sqrt{2d}}
(\sum_{j=1}^{d}|j,j,j\rangle+\sum_{j=1}^{d-1}|j,j+1,d+j\rangle
+|d,1,2d \rangle)$. In addition, an example of the latter type is
constructed by the rule of monoid later. Thus, we can confirm the
tightness of the above inequalities for $\cS_{RNN}$. One can
similarly show the non-emptyness of subsets $\cS_{RRR}$
and $\cS_{RRN}$. Therefore we have justified the classification in
Table \ref{tab:tripartite}. That is, up to permutation of parties
there are 18 different non-empty subsets of tripartite pure states in
terms of the reduced states.



\begin{widetext}

\begin{table}
 \caption{\label{tab:tripartite}
 Classification of tripartite states
 $\ket{\ps}_{ABC}$ in terms of the bipartite reduced states.
 The table contains neither the classes generated from the
 permutation of parties, and nor the subset $\cS_{NNN}$ since for which there is no
 fixed relation between the tensor rank $\rk(\ps)$ and
 {\em local ranks} $d_A(\Psi)$, $d_B(\Psi)$,
 and $d_C(\Psi)$. They are simplified to $r$, $d_A$, $d_B$, and $d_C$ when there is no confusion. All expressions
 are up to local unitaries and all sums run from 1 to $r$.
 In the subset $\cS_{SNN}$, the linearly independent states $\ket{c_i}$ are the support of space $\cH_C$.
    }
\centering
  \begin{tabular}{ | c| c| c| c| c|c| c| c |}
    \hline
    $\cS_{X_{AB}X_{BC}X_{CA}}$
    & $\cS_{SSS}$
    & $\cS_{SSN}$
    & $\cS_{SNN}$
    & $\cS_{PNN}$
    & $\cS_{RRR}$
    & $\cS_{RRN}$
    & $\cS_{RNN}$
    \\
    \hline
    \begin{tabular}{c}
    expression of
    \\
    $\ket{\ps}_{ABC}$
    \end{tabular}
    & $\displaystyle \small \sum_i \sqrt{p_i} \ket{iii}$
    & $\displaystyle \small \sum_i \sqrt{p_i} \ket{i,b_i,i}$
    & $\displaystyle \small \sum_i \sqrt{p_i} \ket{a_i,b_i,c_i}$
    & $\displaystyle \small \sum_i \sqrt{p_i} \ket{a_i,b_i,c_i}$
    & $\displaystyle \small \sum_i \sqrt{p_i} \ket{a_i,b_i,c_i}$
    & $\displaystyle \small \sum_i \sqrt{p_i} \ket{a_i,b_i,c_i}$
    & $\displaystyle \small \sum_i \sqrt{p_i} \ket{a_i,b_i,c_i}$
    \\
    \hline
    \begin{tabular}{c}
    tensor rank
    \\and
    \\local ranks
    \end{tabular}
    &
    \begin{tabular}{c}
    $r=d_A=$\\
    $d_B=d_C$
    \end{tabular}
    &
    \begin{tabular}{c}
    $r=d_A=$\\
    $d_C \ge d_B$
    \end{tabular}
    &
    \begin{tabular}{c}
    $r=$\\
    $d_C \ge d_A, d_B$
    \end{tabular}
    &
    \begin{tabular}{c}
    $r \ge$\\
    $ d_C > d_A, d_B$
    \end{tabular}
    &
    \begin{tabular}{c}
    $r >$\\
    $d_C = d_B = d_A$
    \end{tabular}
    &
    \begin{tabular}{c}
    $r >$\\
    $d_C = d_A \ge d_B$
    \end{tabular}
    &
    \begin{tabular}{c}
    $r \ge d_C \ge d_A, d_B$\\
    $r > d_A, d_B$
    \end{tabular}
    \\
    \hline
  \end{tabular}
\end{table}
\end{widetext}

\medskip
\noindent \textit{Comparison to SLOCC classification}. We know that
there are much efforts towards the classification of multipartite
state by invertible SLOCC \cite{dvc00,cds08,ccd10}. Hence, it is
necessary to clarify the relation between this method and the
classification by reduced states in Table \ref{tab:tripartite}. When
we adopt the former way we have no clear characterization to the
hierarchy of bipartite entanglement between the involved parties,
i.e., the structure of reduced states becomes messy under SLOCC. Our
classification resolves this drawback. Another potential advantage
of our idea is that we can apply the known fruitful results of
bipartite entanglement, such as the hierarchy of entanglement to
further study the classification problem.

Here we explicitly exemplify that the invertible SLOCC only
partially preserves the classification in Table
\ref{tab:tripartite}. We focus on the orbit $\cO_{r=d_A=d_B=d_C}:=
\{ |\Psi\rangle | \rk(\Psi)= d_A(\Psi)=d_B(\Psi)=d_C(\Psi) \}$,
which has non-empty intersection with the subsets $\cS_{SSS}$,
$\cS_{SSN}$, and $\cS_{SNN}$. Further, since the subset $\cS_{NNN}$
contains the state $|\Psi_a\rangle$, it also has non-empty
intersection with $\cO_{r=d_A=d_B=d_C}$. Since all state in
$\cO_{r=d_A=d_B=d_C}$ can be converted to GHZ state by invertible
SLOCC, it does not preserve the classification by reduced states.
However, the subsets $\cS_{RRN}$ and $\cS_{RRR}$ are not mixed with
$\cS_{SSS}$, $\cS_{SSN}$, and $\cS_{SNN}$ by invertible SLOCC.

\medskip
\noindent \textit{Monoid structure}. To get a further understanding
of Table \ref{tab:tripartite} from the algebraic viewpoint, 
we define the direct sum for subsets by
$\cS_{X_{AB}X_{BC}X_{CA}}\cS_{Y_{AB}Y_{BC}Y_{CA}}:=
\cS_{\max\{X_{AB},Y_{AB}\}
\max\{X_{BC},Y_{BC}\}\max\{X_{CA},Y_{CA}\}}$, where $\max\{X,Y\}$ is
the larger one between $X$ and $Y$ in the order $S \le P \le R  \le
N$. Therefore when $\ket{\Psi_1} \in {\cal S}_{X_{AB}X_{BC}X_{CA}}$
and $\ket{\Psi_2} \in {\cal S}_{Y_{AB}Y_{BC}Y_{CA}}$, the state
$\ket{\Psi_1\cdot \Psi_2} \in \cS_{\max\{X_{AB},Y_{AB}\}
\max\{X_{BC},Y_{BC}\}\max\{X_{CA},Y_{CA}\}}$. This product is
commutative and in the direct sum, the subset $\cS_{SSS}$ is the
unit element but no inverse element exists. So the family of
non-empty sets $\cS_{X_{AB}X_{BC}X_{CA}}$ is an abelian monoid
concerning the direct sum.

The above analysis provides a systematic method to produce the
subsets in the monoid. Generally we have
$\cS_{SNN}=\cS_{SSN}\cS_{SNS}, \cS_{RRN} = \cS_{RRR}\cS_{SSN},
\cS_{RNN}=\cS_{RRR}\cS_{PNN}$, and $\cS_{NNN} = \cS_{PNN}\cS_{NSS}$.
So it is sufficient to check the non-emptyness of subsets
$\cS_{SSS}, \cS_{SSN}, \cS_{PNN}, \cS_{RRR}$. By Table
\ref{tab:tripartite}, the first three subsets exist and the
symmetric state $
\frac{1}{\sqrt{2r}}(|312\rangle+|123\rangle+|231\rangle+|213\rangle+|132\rangle+|321\rangle)
+\frac{1}{\sqrt{r}} \sum_{j=4}^r|jjj\rangle$ indicates the
non-emptyness of $ \cS_{RRR}$ for $r \ge 3$. So we have verified the
non-emptyness of eight subsets in the monoid. In particular, if we
choose $\ket{\Psi_1} \in {\cal S}_{SNN}$ and $\ket{\Psi_2} \in {\cal
S}_{RRR}$ and both have $d_A=d_B=d_C $. Then the state
$\ket{\Psi_1\cdot \Psi_2}$ verifies the non-emptiness of the subset
${\cal S}_{RNN}$. Similarly, we can construct an example with $r
> d_A=d_C> d_B$ in ${\cal S}_{RNN}$ by the condition
$d_A(\Psi_1)=d_{C}(\Psi_1) > d_{B}(\Psi_1) $.

\medskip
\noindent \textit{Proof of Theorem \ref{thm:sixconditions}}. We
propose the preliminary lemma.
\begin{lemma}
\label{le:sep} Consider a tripartite state $\ket{\Psi_{ABC}}$ with a separable reduced
state $\rho_{BC}$. When $\rho_{AB}$ satisfies the condition {\bf
(6')}, it also satisfies the condition {\bf (1)}.
\end{lemma}
\begin{proof}
Due to separability, $\rho_{BC}$ can be written by $\rho_{BC}=\sum_{i}p_i |\phi_i^B,\phi_i^C\rangle \langle
\phi_i^B,\phi_i^C|$.
We introduce the new system ${\cal H}_D$ with the orthogonal basis $e_i^D$
and the tripartite extension $\rho_{BCD} :=
\sum_{i}p_i |\phi_i^B,\phi_i^C,e_i^D\rangle \langle
\phi_i^B,\phi_i^C,e_i^D|$.
The monotonicity of the relative entropy $D(\r||\s):=\tr (\r \log \r - \r \log\s)$
and Condition {\bf (6')} imply that
\begin{align*}
 0&=
 H(\rho_{C})-H(\rho_{BC})
 = D(\rho_{BC}\| I_B  \ox \rho_C )
 \nonumber\\
 &\le
 D(\rho_{BCD}\| I_B \ox \rho_{CD} )
 =\sum_{i}p_i (\log p_i -\log p_i)
=0.
\end{align*}
So the equality holds in the above inequality. According to Petz's
condition \cite{Petz}, the channel $\Lambda_C: \cH_C \mapsto \cH_C
\otimes \cH_D$ with the form $\Lambda_C(\sigma):=
\rho_{CD}^{1/2}((\rho_{C}^{-1/2}\sigma \rho_{C}^{-1/2}) \otimes
I_D)\rho_{CD}^{1/2} $ satisfies $ id_{B} \ox \Lambda_C
(\rho_{BC})=\rho_{BCD} $. We introduce the system $\cH_E$ as the
environment system of $\Lambda_C$ and the isometry
$U:\cH_C\mapsto\cH_C \otimes \cH_D \otimes \cH_E $ as the
Stinespring extension of $\Lambda_C$. So the state
$|\Phi_{ABCDE}\rangle:=I_{AB} \ox U\ket{\Psi_{ABC}}$ satisfies
$\rho_{BCD}=\Tr_{AE} \proj{\Phi_{ABCDE}}$. By using an orthogonal
basis $\{e^{AE}_i\}$ on $\cH_A \otimes \cH_E$ we can write up the
state $|\Phi_{ABCDE}\rangle =\sum_{i}\sqrt{p_i}
|\phi_i^B,\phi_i^C,e_i^D,e_i^{AE}\rangle $. Therefore $\r_{AB}$ is
separable. This completes the proof.
\end{proof}
Due to Lemma \ref{le:sep} and the equivalence of conditions {\bf
(5')} and {\bf (6')}, it suffices to show that when $\rho_{BC}$ is
PPT and $\rho_{AB}$ satisfies {\bf (5')}, $\rho_{BC}$ is separable.
From {\bf (5')} for $\rho_{AB}$, it holds that $\rank
\rho_{BC}=d_A=\rank \rho_{AB}=d_C$. So $\rho_{BC}$ is separable by
Lemma \ref{le:rankAB=rankA,B}, and we have Theorem
\ref{thm:sixconditions}.

\medskip
\noindent \textit{Conclusions}. We have proposed the converse
monogamy of entanglement such that when Alice and Bob are weakly
entangled, then either of them is generally strongly entangled with
the third party. We believe that the converse monogamy of
entanglement is an essential quantum mechanical feature and it
promises a wide application in deciding separability, entanglement
distillation and quantum cryptography. Our result presents two main
open questions: First, can we propose a concrete quantum scheme by
the converse monogamy of entanglement? Such a scheme will indicate a
new essential difference between the classical and quantum rules,
just like that from quantum cloning \cite{wz82} and the negative
conditional entropy \cite{how05}. Second, different from the
monogamy of entanglement which relies on the specific entanglement
measures \cite{KW}, the converse monogamy of entanglement only
relies on the strength of entanglement. So can we get a better
understanding by adding other criteria on the strength of
entanglement such as the non-distillability ?

We also have shown tripartite pure states can be sorted into 18
subsets and they form an abelian monoid. It exhibits a more
canonical and clear algebraic structure of tripartite system
compared to the conventional SLOCC classification \cite{dvc00}. More
efforts from both physics and mathematics are required to understand
such structure.

\medskip

\acknowledgments

We thank Fernando Brandao for helpful discussion on the reduction
criterion and Andreas Winter for reading through the paper. The CQT
is funded by the Singapore MoE and the NRF as part of the Research
Centres of Excellence programme. MH is partially supported by a MEXT
Grant-in-Aid for Young Scientists (A) No. 20686026.

\end{document}